\documentstyle[12pt]{article}
\def\L{{\cal L}}

\def\V{{\cal V}}

\def\p{\partial}

\def\L{{\cal L}}

\def\NPB #1#2#3{Nucl. Phys. {\bf B#1#2#3}}

\def\Der#1#2{{d#1\over d#2}}

\begin{document}
\title{Batalin-Vilkoviski master equation\\ and\\ absence of anomalies
in string field theory}
\author{\hspace{-.7in}
     \hspace{-.4in}   Jos\'e Gaite\hspace{1.9in}  Ram Brustein\\
   \hspace{-.3in}  Departamento de F\'{\i}sica\hspace{.7in}
     Department of Physics\\
\ Universidad de Salamanca\hspace{.5in} University of Pennsylvania\\
   \hspace{-.23in} Salamanca 37008. Spain\hspace{.7in} Philadelphia, PA
19104
                }
\date{ }
\maketitle
\vspace{-4in}\rightline{UPR-561-TH/Salamanca FTUS-9307}
\vspace{4.2in}

\begin{abstract}
We study the Batalin-Vilkovisky master equation for both open
and closed string field theory with special attention to anomalies.
Open string field theory is anomaly free once the  minimal
coupling to closed strings induced by loop amplitudes is considered.
In closed string field theory the full-fledged  master equation has to be
solved order by order in perturbation theory.
The existence of a solution implies the absence of anomaly.
We briefly discuss the relation of the iterative
process of solution to  methods used in the first quantized
formalism and comment on some possible non-perturbative corrections.
\end{abstract}

\section{Introduction}

Recent developments in the formulation of string field theory
\cite{Zwiebach},\cite{Witten}
rely on the powerful formalism to quantize gauge theories introduced by
BRST and extended by Batalin and Vilkovisky (BV).
For a recent review of the BRST/BV formalism see \cite{Henneaux}.
Its crucial property is  that it allows to consider gauge theories that
have an open algebra or are reducible. This makes it convenient for
string theory.

     The BV formalism introduces new fields, called antifields and
expresses the BRST invariance in a compact way by a master
equation. The solutions of this equation are all possible gauge
theories in a definite configuration space. In SFT one looks
for an action (function of the string field) that, on the one
hand, reproduces the first quantized string amplitudes and, on the
other, once it is second quantized must have a solution
corresponding to the correct physical theory.

    Unlike the situation in Witten's open string field theory (OSFT), which
has cubic action \cite{osft}, Sonoda and Zwiebach
\cite{Sonoda-Zwiebach}
have shown that for closed string field theory (CSFT) the condition of
modular invariance, necessary to reproduce string amplitudes, can be
 expressed by an equation, called the geometric equation.
It is solved iteratively and the
solution involves an infinite number of vertices at tree  level as well as
an infinite number of ``counterterms" for loop amplitudes.

The BV master equation was introduced by Hata \cite{Hata} in cubic CSFT
to cure its unitarity problems. He used an analogy with the path integral
quantization of the Non-Linear Sigma Model (NLSM). In this model it is
possible to show that if one uses the naive measure in field space the
theory is not unitary. The reason is that the naive path integral measure
is not invariant under the nonlinear symmetry and it is necessary to
add a correction term to restore invariance and unitarity. In SFT one
deals with BRST  symmetry, an elaborate non-linear transformation.
Nevertheless, Hata was able to show that
the BV master equation can be solved at one-loop order in a similar way
to the NLSM case and, with more effort, to higher orders. He further
proved the resulting quantum action to be unitary  up to three-loop order.

The similarity of the geometric equation to the perturbative expansion of
the BV master equation was further clarified by  Kaku \cite{Kaku},
showing a
closer relation between the lack of modular and gauge invariance, which
he called anomaly. He did not use though the BRST formalism nor
attempted to connect with either Sonoda-Zwiebach or Hata's work.

     In this paper we  make explicit the relation
between modular and BRST invariance with special concern for the
measure problem. We base our discussion on the BV master equation,
\begin{equation}
\frac{1}{2} (S,S)-i\hbar\Delta S=0
\end{equation}
which we consider the fundamental equation of SFT.
The relevant term for us is $\Delta S$, which expresses the
failure of BRST (or modular) invariance of the path integral
measure. Nevertheless, it is possible to find a local functional
(counterterm) such that its BRST variation compensates for
$\Delta S$. Hence, strictly speaking, there is no anomaly.
Following Hata, we shall appeal to the NLSM to see that the
basic idea is very simple. However, in CSFT we encounter an important
difference. The BRST transformation is also altered by the
counterterm. The new BRST transformation is again ``anomalous"
at two-loop order and has to be corrected, and so on ad infinitum. This
recursive process represents the iterative solution of the BV master
equation in the loop expansion. In this process, unlike ordinary
iterative schemes the equation as well as the solution change in each
order.

   We begin by reviewing the BRST and BV formalisms with special
attention to the role of anomalies \cite{Troost et al}. To set the scene
for the use of the master equation in SFT and for its own sake as well
we regard its role in OSFT. Here one achieves a complete theory
with just the cubic Witten's action plus a series of terms coupling an
increasing number of open strings to one closed string. It is not
necessary to introduce the master equation though it is convenient
(see \cite{Thorn} for an introduction).
The calculation of $\Delta S$ follows a line analogous to the Yang-Mills
case and must yield $\Delta S=0$ once the divergence is cured by the
introduction of the closed-string terms mentioned above. From the
alternative Riemman-surface point of view, this calculation is not
necessary since there is a fair amount of evidence for the conjecture
that Witten OSFT exactly covers moduli space \cite{Giddings et al,Sam}.
We proceed to CSFT; we briefly comment why there is no classical
polynomial solution to the master equation before we begin with
quantum corrections. We show explicitly how to obtain the one and
two-loop counterterms. We  discuss  gauge
invariance in SFT in comparison with field theory. We end with a brief
discussion of the relevance of our analysis to the first quantized
formalism and some issues regarding non-perturbative effects.

\section {Anomalies in the BRST formalism}

Anomalies can be understood as the non-invariance of the path integral
measure under gauge transformations. Therefore, one should expect that
in the gauge-fixed path integral they manifest themselves as
lack of invariance
under the BRST transformation; or in the BV formulation as the
impossibility to fulfill the master equation. Recent work shows
that this is indeed the case \cite{Troost et al}.

Nevertheless, it is convenient to compare first the BRST anomaly
to a similar problem in a theory that is not gauge, the NLSM,
following \cite{Hata}. This is because the peculiar behavior of
the measure under the BRST transformation $\delta_B$ is mainly due to
its nonlinear character.

The quantum NLSM was found not to be invariant under the
classical symmetry and, as a consequence, unitarity would be
lost. The problem could be expressed as the necessity to add a
term to the classical action to cancel the unwanted counterterms
that break the symmetry. This can be traced back to  the non-gaussian
integral over momenta in the functional integral. Another interpretation
of this
term, more suitable to our purposes, is as the function that converts
the naive path  integral measure into that invariant under the nonlinear
transformation (Haar measure).

The relevant transformations for a definite NLSM are the
isometries (Killing vectors) of the defining manifold
\begin{equation}
\delta_f\phi^i = f^i(\phi),
\end{equation}
satisfying
\begin{equation}
f^k_{,i}g_{kj} + f^k_{,j}g_{ik} + f^k g_{ij,k} = 0.
\end{equation}
Multiplying by $g^{ij}$ we obtain
\begin{equation}
2f^i_{,i} + f^k g^{ij} g_{ij,k} = 2f^i_{,i} + f^k \partial_klng = 0.
\end{equation}
where $g=det g_{ij}$.
Recalling the change of the measure under diffeomorphisms
\begin{equation}
\delta_f lnd\phi = \frac{\partial}{\partial\phi^i}
\delta_f\phi^i = f^i_{,i},
\end{equation}
we obtain
\begin{equation}
\delta_f [lnd\phi + (1/2)lng] = \delta_f [lnd\phi +
(1/2)trlng_{ij}] = 0.
\end{equation}
This correction is local as a functional, which implies the
presence of $\delta^D(0)$, yielding as quantum lagrangian
\begin{equation}
\L_q = \L - i(\hbar/2) \delta^D(0)trlng_{ij}(\phi).  \label{Lq}
\end{equation}
We have corrected the non-invariance of the quantum theory by a local
modification of the
lagrangian and hence there is no actual anomaly. Alternatively, the
measure can be corrected
$\prod_i[d\phi_i]\rightarrow\prod_i[d\phi_i]\sqrt{g}$,
leaving the lagrangian unchanged

BRST is usually a nonlinear transformation and we  expect
that the naive  path integral measure is not invariant under it,
\begin{equation}
\delta_B ln[d\phi] = \frac{\delta}{\delta\phi^i} \delta_B\phi^i
\end{equation}
will generally be some non-null functional. The BV formalism
generalizes BRST with the introduction of antifields,$\phi^*_i$, and the
classical action, $S(\phi^i,\phi^*_i)$, satisfies
\begin{equation}
\delta_B\phi^i = \frac{\delta S}{\delta\phi^*_i},
\end{equation}
and therefore
\begin{equation}
\delta_B ln[d\phi] = \frac{\delta^2 S}{\delta\phi^i\delta\phi^*_i}
\equiv \Delta S.
\end{equation}
Hence, a sufficient condition for the absence of anomalies is
\begin{equation}
\Delta S = 0.
\end{equation}
However, $\Delta S = 0$  is not a necessary condition.
As long as there exists a local functional\footnote
{The dependence on antifields is not shown, since the gauge has to be
eventually fixed and they are then removed.}
$M_1(\phi)$ such that
\begin{equation}
\Delta S = -i\delta_BM_1(\phi),   \label{noano}
\end{equation}
we can absorb the BRST variation of the measure in that of the
action and the path integral as a whole will be invariant,
\begin{equation}
\delta_B [ln[d\phi] + i(S + M_1)] = 0.
\end{equation}
We can write (\ref{noano}) with the use of the antibracket as
\begin{equation}
\Delta S = (M_1,S).   \label{noano1}
\end{equation}
However, the addition of $M_1$ to the classical action modifies the
BRST transformation to the quantum one-loop BRST transformation
\begin{equation}
\delta_B^q\alpha = (\alpha,S+M_1),   \label{BRST1}
\end{equation}
which in turn implies that at two-loop order
\begin{equation}
\delta_B^q [ln[d\phi] + i(S + M_1)] = \Delta M_1 + i{1\over 2}
(M_1,M_1),
\end{equation}
using the previous equations. In order not to have anomaly at
this order, there must exist a function $M_2(\phi)$ that
accounts
for this variation,
\begin{equation}
\Delta M_1 + i{1\over 2}(M_1,M_1) = -i\delta_B^q M_2(\phi),
\end{equation}
and the whole argument repeats itself.

Proceeding this
construction, we get the loop expansion of the full BV master
equation, which expresses in a compact form the invariance of the
path integral under the BRST transformation, or equivalently the
no anomaly condition:
\begin{eqnarray}
2i\hbar\Delta W - (W,W) = 0, \label{noan}\\
W = S + \hbar M_1 + \hbar^2 M_2 + \cdots. \nonumber
\end{eqnarray}
If this equation were to be violated we would definitely have an anomaly.
This violation usually manifest itself in perturbation theory
as impossibility to find the functions $M_i$ but it could also appear in a
non-perturbative manner, for example, as a failure of the
perturbative series for $W$ to converge. It has been proposed that
this problem might arise in string theory and we shall comment on it
below.
The analysis of the NLSM suggests that it may be simpler to correct the
measure than to correct the lagrangian. The final result is, however, the
same.

\section{Quantization of OSFT.}

The question of BRST invariance can be divided into a classical part,
namely to
find an $S$ that fulfills $(S,S) = 0$, and a quantum part that begins
with checking whether $S$ satisfies $\Delta S = 0$. If it does,
the full quantum theory is BRST invariant and the iterative process ends.
The actual calculation of $\Delta S$ involves divergences
and requires the introduction of a regulator.
In field theory there are several regularization
methods available, which we separate in two, dimensional and
cutoff-like regularization. The former is not convenient since it
hides some divergences (e.g. quadratic), for example, yielding a vanishing
quantum contribution in Eq.(\ref{Lq}). The second is thus more
adequate; in particular, a variant of the Pauli-Villars scheme
that has been proposed in \cite{Troost et al}. This method has
been applied to Yang-Mills (YM) gauge theory \cite{De Jonghe et al};
the value of $\Delta S$ to be regularized is
\[\delta^D(0)\: f^a_{ba} c^b = \infty \times 0.\]
It is nonvanishing in a general gauge but its value can be
absorbed by local counterterms, not surprisingly since pure YM is
a renormalizable anomaly-free theory.

In Witten's OSFT the computation of $\Delta S$ was undertaken by Thorn
but his result is inconclusive \cite{Thorn}. It has been later argued by
Kaku that it must be zero\footnote{In fact, Kaku considers instead the
jacobian of the gauge transformation but its vanishing is equivalent to
the vanishing of $\Delta S$}  \cite{Kaku}. The argument relies on an
analogy with YM: The gauge group of OSFT can be formulated in a similar
form to an ordinary gauge group and thus its functional structure
constants are antisymmetric as well \cite{Kaku_book}.
The value of $\Delta S$ is
\[\prod_\sigma \delta[X(\sigma)-X(\sigma)]\: f^i_{ji} \Phi^j. \]
It vanishes provided that one disregards the divergence. However, it can
occur as in YM, namely, that a careful computation does not yield zero.
We actually know that this one-loop divergence can be associated with
coupling to a closed string. Therefore, we must take into account the
interaction of a closed string with an arbitrary number of open strings
which is originated in this way,
\begin{equation}
S_{int-oc} = \int \Psi (\Phi + \Phi^2 + \Phi^3 + \cdots).  \label{oci}
\end{equation}
The OSFT that includes these interactions satisfies the classical master
equation \cite{Zwiebach-OSFT}.

Similar conclusions can be reached from the Rieman-surface point of
view, seeking covering of moduli space. There is sufficient evidence by
now that amplitudes formed with Witten vertices correctly fill the
relevant moduli spaces, even there is no general mathematical proof
\cite{Giddings et al,Sam}.
Thus the full quantum theory is modular
(hence BRST) invariant and Witten's
cubic action, including the closed string interaction (\ref{oci}), does not
need further addition of quantum counterterms.

\section{Quantization of CSFT}

It has long been known that the SFT program and, in particular,
Witten's cubic action encounter problems when adapted to closed
strings \cite{Giddings and Martinec}.
As a partial solution, it was proposed to introduce new
vertices,  originating from a nonpolynomial action
\cite{Kugo et al,Kaku2,Saadi-Zwiebach}
that constitutes a modular invariant theory at tree level.
This expansion is analogous to the infinite expansion of the Einstein-
Hilbert action $\sqrt{g}R$ around a particular background. In fact, the
mode expansion of the theory contains the expansion of $\sqrt{g}R$.
It is therefore not surprising that an infinite number of vertices is
necessary.
Loop amplitudes were later analysed
\cite{Sonoda-Zwiebach,Hua-Kaku}. From the requirement of single
covering of moduli space, Sonoda
and Zwiebach concluded with the necessity to add an infinite
number of vertices with increasing number of external legs for
each genus. In so doing they arrived to a geometric equation as
a consistency condition,

%
\vspace{47mm}
\centerline{Figure 1. Pictorial description of the geometric equation.}
\centerline{$a$ is the maximal allowed value of the sewing parameter
$t$.}
\vspace{5mm}

This equation can be cast in a different form \cite{Brustein-De Alwis},
reinterpreting it as a  Wegner-Polchinski renormalization
group equation for CSFT,
\begin{equation}
a\sum_{N\geq2}\p\V_{G,N} = a\Der {S_{int}} a =
\frac{\delta S_{int}}{\delta \Psi}
\frac{\delta S_{int}}{\delta \Psi} -
\frac{\delta^2 S_{int}}{\delta \Psi^2}.
\end{equation}
The left hand side of this equation is the variation of the
interaction action due to an infinitesimal change of a stub boundary.
Alternatively, it can be regarded as the derivative w.r.t. the
maximal lenght of the sewing parameter, which acts as a short-distance
world-sheet cutoff  in this picture. Interestingly, it can be written as
the linear (first-quantized) BRST variation of the interaction,
\begin{equation}
a\Der {S_{int}} a = QS_{int} =
\frac{\delta S_{int}}{\delta \Psi} Q\Psi =
\frac{\delta S_{int}}{\delta \Psi}
\frac{\delta S_2}{\delta \Psi}.
\end{equation}
Hence, the geometric equation simplifies to
OA\begin{equation}
\frac{\delta S}{\delta \Psi}
\frac{\delta S}{\delta \Psi} -
\frac{\delta^2 S}{\delta \Psi^2} = 0.  \label{BVH}
\end{equation}
This is Hata's version of the BV equation for CSFT, as was
already realized in \cite{Brustein-De Alwis,Zwiebach3}.
It embodies BRST invariance in just the same way as modular
invariance in the original SZ picture, showing the equivalence
of both concepts.

However, the one-loop calculation takes very different forms
whether it is made in the BRST or Riemman-surface formalisms. In
the latter the calculation is performed in ref.
\cite{Zemba-Zwiebach} (see also \cite{Hua-Kaku}) and the divergence
appears because of the necessary
inclusion of an infinite number of torus modular regions.
BRST loop calculations were made by Hata for the purely cubic
theory \cite{Hata}.

Hata's one-loop computation resembles somewhat the one in
standard QFT, though the divergence is more severe due to the
propagation of an infinite number of modes round the loop. The
essential part is the (one-loop) one-leg counterterm $M_{1,1}$,
solution of the corresponding part of Eq. (\ref{BVH}),
\begin{equation}
M^i_{1,1} Q\Psi_i \equiv  QM_{1,1} =
\frac{\delta^2 S_3}{\delta\Psi_i \delta\Psi_i}, \label{BVH1,1}
\end{equation}
which is the simplest version of (\ref{noano}). Multiplying by the
propagator $Q^{-1}$,
\begin{equation}
M_{1,1} =
Q^{-1}\frac{\delta^2 S_3}{\delta\Psi_i \delta\Psi_i} =
Tr[Q^{-1}\Psi*].     \label{M1,1}
\end{equation}
The divergence is exposed by explicitly writing  the propagator as a sum
of modes. The trace is equivalent to performing the integration
over momenta round the loop.

The two-leg equation
\begin{equation}
M^i_{1,1} \frac{\delta S_3}{\delta\Psi_i} + M^i_{1,2} Q\Psi_i =
\frac{\delta^2 S_4}{\delta\Psi_i \delta\Psi_i}, \label{BVH1,2}
\end{equation}
differs from Hata's in the presence of a term on the right-hand
side coming from the quartic vertex, which he neglects. It can
be solved  to yield
\begin{equation}
M_{1,2} = - M^i_{1,1} Q^{-1} \frac{\delta S_3}{\delta\Psi_i} +
Q^{-1}\frac{\delta^2 S_4}{\delta\Psi_i \delta\Psi_i}.
\label{M1,2}
\end{equation}
It contains a quantum correction to the kinetic term.

We can pursue this procedure for greater number of legs but the
equations and hence their solutions become increasingly complex.
Restricting to the cubic vertex only, Hata gives a closed
expression for the one-loop action much in the style of standard QFT
\cite{Hata}. He also proceeds to higher loops.
Unfortunately, his expressions are formal because they are divergent and
he does not introduce any regularization. Therefore, one cannot read off
the
values of the $M_n$ from them. For the same reason, it is impossible to
read off $M_n$ from Kaku's computations \cite{Kaku}.

To see concrete solutions we have to appeal to the work done in the
Riemman-surface formalism. The vertices $\V_{G,N}$, solution of the
Sonoda-Zwiebach equation, are to be identified with $M_{n,N}$ ($n=G$).
However, the corresponding calculations, comprising analysis of the
divergences, have been carried out to a
much lesser extent; namely, only one-loop calculations are
available and only for one or two punctures \cite{Saadi,Hua-Kaku}.

A comparison with the first quantized formalism is in order here.
In \cite{Brustein-De Alwis} and in \cite{Brustein-Roland} the relation
between the second quantized formalism based on solving the classical
BV equation and the first quantized formalism based on solving
the conditions of conformal invariance was explained (This
was done in a limited framework, involving only the massless modes
and in the limit $a\rightarrow 0$).
The main result was that a solution to the conformal invariance
conditions automatically solves the classical BV equation.
The situation is  different when the full quantum BV equation
is considered. In the limit $a\rightarrow 0$, the dependence of the linear
term $\Delta S$ on $a$ is completely different than the one coming from
$(S,S)$ and therefore both terms cannot be treated on the same footing.
 Because of that the contribution coming from the ``non-dividing pinch"
\cite{bda}, was largely ignored. Note that the divergences
coming from the term $\Delta S$ are not those known
as Fischler-Susskind  divergences \cite{Brustein-Roland}.
As realized in \cite{Marcus} unitarity dictates certain analytic
continuation that turn the divergence into an imaginary part. From
the BV equation for SFT, it is clear, however, that both terms have to be
included. In fact, it has long been conjectured \cite{bda},\cite{Shenkcon}
that these terms are responsible for the anomalous $(2G)!$ growth of
 large order perturbation series in string theory \cite{Shenker}.
A proof of this conjecture should be helpful in deciding whether there
are
any non-perturbative obstructions (``anomalies") to finding a complete
solution.

\section{Conclusions and Discussion}

We have studied anomalies for both OSFT and CSFT in the BV
formalism, spelling out the equivalence of BV master equation
with Sonoda-Zwiebach geometric equation. We have shown how to solve
this equation to obtain the quantum correction to the measure. The
concrete solution is not very illuminating and the
important point is whether it can be solved, which implies the absence
of anomalies.
An analogy with first-quantized string theory is appropriate: In
this case, the BRST charge $Q$ is nilpotent except for the
appearance of the conformal anomaly. In the second-quantized
theory one pursues as well $\delta_B^2 = 0$. We know that this
condition is equivalent to the master equation $(S,S)=0$. This
equation is solved in OSFT by Witten's interaction. In CSFT the
classical master equation is complemented by quantum corrections,
hence $\delta_B^2 \neq 0$. Nevertheless, there is no anomaly
provided we can solve the quantum BV master equation, thus
finding a nilpotent quantum BRST transformation.

With regard to anomalies our answer is that they seem to be absent
on the formal level of calculations presented here.
We would like to add however some words of caution. It is obvious that
the situation in CSFT is complicated, since the infinity of vertices in
the non-polynomial action is corrected at each loop order.
It is desirable to further investigate the issue and to make
the absence of anomalies manifest. One way of achieving this goal would
be to find a gauge in which $\Delta S=0$. We know that the solution of
the BV master equation is gauge dependent and we might expect
that for some yet unknown gauge the equation $\Delta S = 0$
would be satisfied. This gauge would be particularly useful
in separating the question of background independence and invariance of
the measure. Another way of making the absence of anomalies manifest
would be to find a string field redefinition for
which the jacobian exactly cancelled $\Delta S$. The covariant
formalism of Schwarz \cite{Schwarz} as presented in
\cite{Hata-Zwiebach}, is useful in stating the problem clearly.
In this formalism this condition can
be expressed as an equation for the measure $\rho$,
$\Delta_\rho^2=0, \ \ \frac{1}{2}(\log\rho,S)+\Delta_1S=0$, where
$\Delta_1$ is the naive measure.

We have mentioned in section 2 the possibility of non-perturbative
violation of Eq.(\ref{noan}), what we should call a non-perturbative
anomaly. It can occur in two related ways:
The first, that the sum does not converge, typically because terms grow
too fast.
The second, that a piece is missing and cannot be reached by perturbation
theory. It would be interesting to obtain from the quantum BV equation
recursion relations that will shed light on this issue.

\section{Acknowledgments}
We gratefully acknowledge discussions with R. Siebelink and B. Zwiebach.
J. G. would like to thank  S. Weinberg for hospitality in the theory group
at the University of texas at Austin, where this work was begun.
The work of R.B. was supported in part by the Department of Energy under
contract No. DOE-AC02-76-ERO-3071.
The work of J.G. was supported in part by CICYT.

\end{document}